%% file: aaamain.tex
\documentclass{article}

\usepackage[backend=biber,sorting=none,style=trad-abbrv,maxbibnames=1]{biblatex}

\usepackage[subpreambles=true]{standalone}
\usepackage{amsmath}
\usepackage{csquotes}
\usepackage{hyperref}
\usepackage[a4paper,margin=1in]{geometry}

\usepackage{xcolor}

\AtEveryBibitem{
    \clearfield{note}
    \clearfield{pagetotal}
    \clearfield{isbn}
    \clearfield{url}
    \clearfield{doi}
    \clearfield{issn}
    \clearfield{urlyear}
    \clearfield{urlmonth}
    \clearfield{eprinttype}
    \clearfield{eprint}
    \clearfield{series}
    \clearfield{volume}
    \clearfield{place}
}

\begin{document}

\input{title.tex}

\input{abstract.tex}

\input{introduction.tex}

\input{method.tex}

\input{results.tex}

\input{discussion.tex}

\input{conclusion.tex}

\input{acknowledgements.tex}

\printbibliography

\end{document}

%% file: title.tex
\maketitle

%% file: abstract.tex
\begin{abstract}
    In a recent paper, algebraic descriptions for all non-relativistic spins were derived by elementary means directly from the Lie algebra $\specialorthogonalliealgebra{3}$, and a connection between spin and the geometry of Euclidean three-space was drawn. However, the details of this relationship and the extent to which it can be developed by elementary means were not expounded. In this paper, we will reveal the geometric content of the spin algebras by realising them within a novel, generalised form of Clifford-like algebra. In so doing, we will demonstrate a natural connection between spin and non-commutative geometry, and discuss the impact of this on the measurement of hypervolumes and on quantum mechanics.
\end{abstract}

%% file: introduction.tex
\section{Introduction}

\subsection{The Spin Algebras \texorpdfstring{$\spinalgebra{s}$}{A(s)}}\label{sec:spin-algebras}

A recent paper\cite{bradshaw} derived real associative algebras that completely describe the spin structure for non-relativistic systems of arbitrary spin. These \enquote{spin algebras} $\spinalgebra{s}$, are derived from the universal enveloping algebra\cite{humphreys} $\usothree$ of the Lie algebra $\sothree$,
\begin{subequations}
\begin{gather}
    \definition{\sothree}{\textup{span}_{\reals}(\set{\generator[1],\generator[2],\generator[3]})}\\
    \sothreelieproductdefinition{a}{b}{c}\label{eqn:so3-lie-product},
\end{gather}
\end{subequations}
\noindent by quotient,
\begin{equation}\label{eqn:spin-algebra-definition}
    \definition{\spinalgebra{s}}{\frac{\usothree}{\ideal{\image{\multipole{2s+1}}}}},
\end{equation}
\noindent where $\ideal{\image{\multipole{k}}}$ is the two-sided ideal generated by the totally-symmetric and contractionless \enquote{multipoles},
\begin{subequations}
\begin{gather}
    \mapdeclaration{\multipole{k}}{\tensorpower{\tsothree}{k}}{\usothree}\\
    \composition{\sothreead[\sothreecasimirelement+k(k+1)]}{\multipole{k}}=0\label{eqn:multipole-eigenobject}\\
    \forall\tau\in S_k,\; \composition{\multipole{k}}{\tau}=\multipole{k}\\
    \forall m\neq n\in \set{1,...,k},\;\sum_{a_m,a_n=1}^{3}\delta_{a_m a_n}\multipole{k}\Big(\bigotimes_{j=1}^{k}\generator[a_j]\Big)=0,
\end{gather}
\end{subequations}
\noindent with $\tsothree$ the tensor algebra of $\sothree$\cite{bourbaki}.

The multipoles are defined recursively in terms of the adjoint action $\sothreead$,
\begin{equation}\label{eqn:adjoint-action}
    \definition{\sothreead[u]}{\mapdefinition{v}{
        \begin{cases}
            uv & u\in\reals\\
            \tensor{u;v}-\tensor{v;u} & u\in\sothree\\
            \sothreead[a]\!\circ\!\sothreead[b](v) & u=\tensor{a;b},
        \end{cases}
    }}
\end{equation}
\noindent the left multiplication $\forall A\in\usothree$,
\begin{equation}\label{eqn:left-multiplication}
    \definition{\sothreeleft[A]}{\mapdefinition{B}{\tensor{A;B}}},
\end{equation}
\noindent and the Casimir element of $\usothree$,
\begin{equation}
    \sothreecasimirelementdefinition{a},
\end{equation}
\noindent as $\forall k\in\integers^{+}$, $\alpha\in\reals$, $v\in\sothree$, $B_{k}\in\tensorpower{\sothree}{k}$,
\begin{equation}
\begin{gathered}\label{eqn:multipole-chain-identity}
    \multipole{0}(\alpha)=\alpha\\
    \multipole{1}(v)=v\\
    \multipole{k+1}(\tensor{v;B_{k}})=\composition{\frac{\composition{\sothreead[\sothreecasimirelement+k(k-1)]}{\sothreead[\sothreecasimirelement+k(k+1)]}}{4(k+1)(2k+1)}}{\sothreeleft[v]}{\multipole{k}(B_{k})}.
\end{gathered}
\end{equation}
\noindent The multipoles are important to the structure of $\usothree$, since $\forall A_{k}\in\usothree$, $\forall k\in\naturals$, for which $\sothreead[\sothreecasimirelement+k(k+1)](A_{k})=0$ may be written as an $\reals[\sothreecasimirelement]$-linear combination of objects from $\image{\multipole{k}}$, and all elements of $\usothree$ are linear combinations of such $A_{k}$. For compactness, let us define $\forall k\in\integers^{+}$,
\begin{equation}
\begin{gathered}
    \definition{\multipoletensor{}}{\multipole{0}(1)}\\
    \definition{\multipoletensor{a_{1}a_{2}...a_{k}}}{\multipole{k}(\tensor{\generator[a_{1}];\generator[a_{2}];...;\generator[a_{k}]})}.
\end{gathered}
\end{equation}
\noindent The spin algebras $\spinalgebra{s}$ are real unital associative algebra of multipoles, $\forall k\in\integers^{+}$,
\begin{equation}
\begin{gathered}
    \spinalgebra{0}\cong\textup{span}_{\reals}(\set{\multipoletensor{}})\\
    \spinalgebra{k}\cong\textup{span}_{\reals}(\set{\multipoletensor{},\multipoletensor{a_{1}},...,\multipoletensor{a_{1}...a_{2k}}}),
\end{gathered}
\end{equation}
\noindent within which,
\begin{equation}
    \sothreecasimirelement=-s(s+1).
\end{equation}

Since all multipoles $\multipoletensor{a_{1}...a_{2k}}$ are algebraic combinations of the $\set{\generator[a]}$, the $\spinalgebra{s}$ encode the spin structures for arbitrary spins $s$ entirely in terms of $\sothree$. The Lie algebra $\sothree$ generates the Lie group $\specialorthogonalgroup{3}$, which is the connected symmetry group of Euclidean three-space. In this way, the $\spinalgebra{s}$ are connected to the geometry of Euclidean three-space, however the extent and consequences of this connection is unclear.

\subsection{A Geometric Realisation of \texorpdfstring{$\sothree$}{so(3)} through Clifford Algebra}\label{sec:clifford-introduction}

To understand the extent of the relationship between the $\spinalgebra{s}$ and the geometry of Euclidean three-space, let us first attempt to understand the underlying geometric content of $\sothree$, with which any geometric account of $\spinalgebra{s}$ must be compatible. Towards this, we will explore the geometric structure of the more general $\sopq$, the Lie algebra of the connected symmetry group $\textup{SO}^{+}(p,q,\reals)$, which preserves the geometry of a $(p{+}q)$-dimensional space with indefinite signature.

Let $\genericmetricspace$ denote such a non-trivial finite-dimensional vector space $\genericspace$ over $\reals$ equipped with a symmetric, non-degenerate, bilinear map $\genericmetricdeclaration$, which we shall follow relativity by referring to as a \enquote{metric}. We may identify the Lie algebra $\sopq$ with the set of all linear maps $A\in\genericspaceendomorphisms$ satisfying, $\forall v,w\in\genericspace$,
\begin{equation}
    \genericmetric[A(v)][B]+\genericmetric[v][A(w)]=0.
\end{equation}
\noindent Such maps are closed under commutators, which serves as the Lie product. It has long been known that $\sopq$ is in bijection\cite{fulton-harris} with $\genericantisymmetrictensors{2}\subset\generictensoralgebra$, the space of second-order antisymmetric tensors on $\genericspace$\cite{bourbaki},
\begin{equation}\label{eqn:antisymmetric-tensors}
    \definition{\genericantisymmetrictensors{2}}{\textup{span}_{\reals}(\set{\tensorwedge{a}{b}\,\vert\,a,b\in\genericspace})},
\end{equation}
\noindent where $\wedge$ is the multilinear, totally antisymmetric, associative \enquote{wedge product}, $\forall k\in\integers^{+}$,$\forall v_{j}\in\genericspace : j\in\set{1,\dots,k}$,
\begin{equation}\label{eqn:n-blades}
    \tensorwedge{v_{1}}{v_{2}}{\dots}{v_{k}}=\frac{1}{k!}\smashoperator{\sum_{\sigma\in S_{k}}}\textup{sgn}(\sigma)\bigotimes_{j=1}^{k}v_{\sigma(j)},
\end{equation}
\noindent with $S_{k}$ is the set of all permutations of $k$ objects, and $\textup{sgn}(\sigma)$ is the sign of the permutation $\sigma$. Explicitly, this bijection may be given, up to a scalar, as $\forall v,w,x\in\genericspace$,
\begin{equation}\label{eqn:sothree-action-generic}
    \mapdefinition{\tensorwedge{v}{w}}{\big(\mapdefinition{x}{\genericmetric[v][x]w-\genericmetric[w][x]v}\big)}.
\end{equation}

This bijection grants us an immediate geometric interpretation for the objects of $\sopq$: linear combinations of planar elements. More generally, the \enquote{$k$-blade}\cite{doran-lasenby} \eqref{eqn:n-blades} can be interpreted as a hypervolume element of dimension $k$. For $\sothree$, a $2$-blade encodes both the plane and angle of the rotation it generates. We refer to an arbitrary element of $\genericantisymmetrictensors{k}$ as a \enquote{$k$-vector}\cite{doran-lasenby} or \enquote{\textit{prefix-}vector} e.g. $2$-vector and bivector are identical. For completeness, we consider $0$-vectors and $0$-blades to be the scalars of $\genericspace$. 

With the objects of $\sopq$ algebraically identified as bivectors $\genericantisymmetrictensors{2}$, we may find their Lie product by constructing the Clifford algebra\cite{crumeyrolle} $\genericcliffordalgebra$:
\begin{equation}\label{eqn:clifford-quotient}
    \genericcliffordalgebra\cong\frac{\generictensoralgebra}{\ideal{\tensor{v;w}+\tensor{w;v}-2\genericmetric[v][w]}}.
\end{equation}
\noindent This quotient reduces all tensors of $\generictensoralgebra$ to linear combinations of $k$-blades. The survival of the $k$-blades in $\generictensoralgebra$ mark them as objects of geometric significance. $\genericcliffordalgebra$ is finite-dimensional, as all $k$-blades with $k>\dim(\genericspace)$ are $0$ by antisymmetry. Since the field of scalars of $\genericspace$ is not of characteristic $2$, this is equivalent to the construction of $\genericcliffordalgebra$ using a quadratic form\cite{crumeyrolle}.

The structure of the Clifford algebra reveals the Lie product between bivectors,
\begin{equation}\label{eqn:clifford-lie-bracket}
\begin{aligned}
    \tensor{(\tensorwedge{a}{b});(\tensorwedge{c}{d})}-\tensor{(\tensorwedge{c}{d});(\tensorwedge{a}{b})}=2\genericmetric[b][c](\tensorwedge{a}{d})-2\genericmetric[b][d](\tensorwedge{a}{c})-2\genericmetric[a][c](\tensorwedge{b}{d})+2\genericmetric[a][d](\tensorwedge{b}{c}),
\end{aligned}
\end{equation}
\noindent turning $\genericantisymmetrictensors{2}$ into a Lie algebra. This Lie product is related to the usual one\cite{weinberg-qft} by a scaling. The Clifford algebra also naturally defines an $\sopq$-action on vectors $\forall a,b,c\in\genericspace$,
\begin{equation}\label{eqn:clifford-bivector-action}
    m(\tensorwedge{a}{b})(c)=\half\big(\tensor{c;(\tensorwedge{a}{b})}-\tensor{(\tensorwedge{a}{b});c}\big)=\genericmetric[a][c]b-\genericmetric[a][b]c,
\end{equation}
\noindent which is identical to \eqref{eqn:sothree-action-generic}. This enables a natural action of the symmetry group $\textup{SO}^{+}(p,q,\reals)$ to be defined algebraically on $\genericcliffordalgebra$\cite{doran-lasenby}.

Restricting our attention to three-dimensional Euclidean space $\threemetricspace$, we may introduce the transformation,
\begin{equation}\label{eqn:bivector-generator-transform-clifford}
    \definition{\generator[p]'}{-\frac{1}{4}\sothreesum*[r]{a,b}\sothreestructureconstants{a}{b}{p}\,\tensorwedge{e_{a}}{e_{b}}},
\end{equation}
\noindent where $\set{e_{1},e_{2},e_{3}}$ are a basis for $\threespace$ satisfying $\threemetric[e_{a}][e_{b}]=\delta_{ab}$. Then, on basis bivectors the Lie product \eqref{eqn:clifford-lie-bracket} becomes,
\begin{equation}
    \commutator{\generator[p]'}{\generator[q]'}=\sothreesum*[r]{r}\sothreestructureconstants{p}{q}{r}\generator[r]',
\end{equation}
\noindent consistent with \eqref{eqn:so3-lie-product}. Thus, we see that $\threecliffordalgebra$ algebraically realises $\sothree$ in a geometrically meaningful way using bivectors.

\subsection{Limitations of the Clifford Algebra Approach}\label{sec:clifford-limitations}

Despite this natural emergence of $\sothree$ within $\threecliffordalgebra$, this realisation is severely limited. To see this, we note that in $\threecliffordalgebra$, we have $\forall a,b,c,d\in\threespace$,
\begin{equation}
    \half\big(\tensor{(\tensorwedge{a}{b});(\tensorwedge{c}{d})}+\tensor{(\tensorwedge{c}{d});(\tensorwedge{a}{b})}\big)=\genericmetric[a][d]\genericmetric[b][c]-\genericmetric[a][c]\genericmetric[b][d].
\end{equation}
\noindent Applying \eqref{eqn:bivector-generator-transform-clifford}, we find in $\threecliffordalgebra$,
\begin{equation}\label{eqn:clifford-symmetric-bivectors}
    \half(\tensor{\generator[p]';\generator[q]'}+\tensor{\generator[q]';\generator[p]'}) = -\frac{1}{4}\delta_{pq},
\end{equation}
\noindent and the Casimir element $\sothreecasimirelement$ of $\sothree$ is,
\begin{equation}\label{eqn:clifford-casimir}
    \definition{\generator'^2}{\sum_{p=1}^{3}\tensor{\generator[p]';\generator[p]'}}=-\frac{3}{4}.
\end{equation}
\noindent Together, \eqref{eqn:clifford-symmetric-bivectors} and \eqref{eqn:clifford-casimir} imply that the spin quadrupole $\multipoletensor{pq}'=0$ in $\threecliffordalgebra$. By the multipole recurrence relationship \eqref{eqn:multipole-chain-identity}, we also conclude that all spin multipoles $\multipoletensor{p_{1}\dots,p_{k}}=0$  for $k>2$. This shows that, unsurprisingly\cite{doran-lasenby}, the unital subalgebra $\set{\reals,\threeantisymmetrictensors{2}}\subset\threecliffordalgebra$ has spin-$\half$ structure, and is algebra isomorphic to $\spinalgebra{\half}$. This is a direct result of the defining algebraic structure \eqref{eqn:clifford-quotient} of $\threecliffordalgebra$. Thus, $\threecliffordalgebra$ cannot support an arbitrary spin structure within it, and cannot be used to explore the geometric content of $\spinalgebra{s}$ for $s\neq\half$.

Finding an algebra which can will be the focus of this paper. In section \ref{sec:method}, we will define a \enquote{Spinless Weak Clifford Algebra} compatible with the structure of an arbitrary $\spinalgebra{s}$. We will present the \enquote{Spin-$s$ Weak Clifford Algebras} derived from these in section \ref{sec:results}, and show they may naturally entail spin-dependence in the measured sizes of hypervolumes. Finally, in section \ref{sec:discussion}, we will discuss the connection between the spin-$s$ Clifford algebras and non-commutative geometries, and contrast these new algebras with other higher-spin models. We will also consider the implications of these algebras for quantum mechanics.

%% file: method.tex
\section{Method}\label{sec:method}

\subsection{Towards a weaker Clifford Algebra}

As we saw in section \ref{sec:clifford-limitations}, the incompatibility of the Clifford algebra with arbitrary spin algebras $\spinalgebra{s}$ is inherent to its algebraic structure. This is unfortunate, since this algebraic structure enabled us to find both a natural Lie algebra action \eqref{eqn:clifford-bivector-action} and the geometric structure of $\sothree$ \eqref{eqn:clifford-lie-bracket}. To study the $\spinalgebra{s}$, we will construct a weaker algebra with both of these features by elementary means. Such an algebra will have no spin structure at all, enabling any $\spinalgebra{s}$ to be embedded within. We expect the Lie product of $\sopq$ to follow from the $\sopq$-action on our algebra, as \eqref{eqn:clifford-lie-bracket} follows from \eqref{eqn:clifford-bivector-action}. Thus, to proceed we require only: an elementary derivation of an $\sopq$-action on vectors; and a determination of how this action must be implemented within an associative algebra.

\subsubsection{Elementary Derivation of the \texorpdfstring{$\sopq$}{so(p,q)}-action on Vectors}\label{sec:derivation-lie-action}

Recall that $\genericspace$ is a finite-dimensional real vector space and that $\genericmetric$ is a non-degenerate, symmetric, bilinear map on $\genericspace$. For any non-null element $a\in\genericspace$, i.e. $\genericmetric[a][a]\neq0$, we may always find a unique direct sum decomposition of $\genericspace$,
\begin{equation}\label{eqn:orthogonal-decomposition}
    \genericspace\cong\textup{span}_{\reals}(\set{a})\oplus W_{a},
\end{equation}
\noindent where $\forall w\in W_{a},\,\genericmetric[a][w]=0$. Specifically, we can write $\forall v\in\genericspace$ uniquely as,
\begin{equation}\label{eqn:orthogonal-decomposition-vector}
    v=\frac{\genericmetric[a][v]}{\genericmetric[a][a]}a+b
\end{equation}
\noindent where $\genericmetric[a][b]=0$. We notice from \eqref{eqn:orthogonal-decomposition-vector} that the decomposition \eqref{eqn:orthogonal-decomposition} is scale-invariant in two ways: $\forall\alpha\in\reals/\set{0}$, $a$ and $a'=\alpha a$ give the same decomposition, as do $\genericmetric$ and $b=\alpha\genericmetric$. This suggests that some of the structure imparted on $\genericspace$ by $\genericmetric$ is independent of scale. Thus, let us explore this structure more easily by accepting scaling of the metric $\genericmetric$ in our arguments. Doing so will enable us to establish results using more mathematically convenient objects.

From \eqref{eqn:orthogonal-decomposition}, we may use $\genericmetric$ to alter the scale of the component of $v\in\genericspace$ parallel to a non-null vector $a\in\genericspace$, $\forall k\in\reals$,
\begin{equation}\label{eqn:component-scale-map}
    \definition{S(k,a)}{\mapdefinition{v}{\genericmetric[a][a]v+(k-1)\genericmetric[a][v]a}},
\end{equation} 
\noindent accepting the overall scaling by $\genericmetric[a][a]$ that occurs. Note that $\forall v,w\in\genericspace$,
\begin{equation}
    \genericmetric[S(k,a)(v)][\,S(k,a)(w)]=\genericmetric^{2}(a,a)\genericmetric[v][w],
\end{equation} 
\noindent precisely when $k_{\pm}=\pm1$. Recognising the $k_{+}$ case as simply an overall scaling, we use the $k_{-}$ solution to define the \enquote{conformal reflection},
\begin{equation}
    \definition{R(a)}{S(-1,a)=\genericmetric[a][a]v-2\genericmetric[a][v]a}.
\end{equation}
\noindent The conformal reflections are a superset of the traditional reflections, but are similarly not closed under composition.

To see this, let us first define a \enquote{\gadjoint} of an endomorphism $A\in\genericspaceendomorphisms$ as an endomorphism $B\in\genericspaceendomorphisms$ such that $\forall v,w\in\genericspace$, 
\begin{equation}
    \genericmetric[A(v)][w]=\genericmetric[v][B(w)].
\end{equation}
\noindent Since $\genericmetric$ is symmetric and non-degenerate, $B$ is unique and its \gadjoint\space is $A$; accordingly, we shall denote the \gadjoint\space of $A$ as $\bar{A}$ with $\bar{\bar{A}}=A$. We also define \enquote{self-\gadjoint} to mean $\bar{A}=A$, \enquote{anti-self-\gadjoint} to mean $\bar{A}=-A$, and two maps $a_{+}$ and $a_{-}$ that respectively yield the self-\gadjoint\space and anti-self-\gadjoint\space parts of an endomorphism $A$,
\begin{equation}
    \definition{a_{\pm}}{\mapdefinition{A}{\half(A\pm\bar{A})}}.
\end{equation}
\noindent
We find all conformal reflections are self-\gadjoint, but the \gadjoint\space of $\composition{R(a)}{R(b)}$ is $\composition{R(b)}{R(a)}$ which is different in general. This indicates that there is structure in the product of two conformal reflections that is not itself a conformal reflection.

To identify this additional structure, we decompose $\composition{R(a)}{R(b)}(v)$, $\forall v\in\genericspace$ into self-\gadjoint\space and anti-self-\gadjoint\space parts $\forall v\in\genericspace$,
\begin{subequations}
\begin{gather}
    a_{-}\big(\composition{R(a)}{R(b)}\big)(v)=-2\genericmetric[a][b]\big(\genericmetric[a][v]b-\genericmetric[b][v]a\big)\label{eqn:conformal-reflection-commutator-proto}\\
    a_{+}\big(\composition{R(a)}{R(b)}\big)(v)=\genericmetric[a][a]\genericmetric[b][b]v+2\genericmetric[a][v]\big(\genericmetric[a][b]b-\genericmetric[b][b]a\big)-2\genericmetric[b][v]\big(\genericmetric[a][a]b-\genericmetric[b][a]a\big).\label{eqn:conformal-reflection-anticommutator-proto}
\end{gather}
\end{subequations}
\noindent Noticing that there is a repeated pattern within \eqref{eqn:conformal-reflection-commutator-proto} and \eqref{eqn:conformal-reflection-anticommutator-proto}, we define $\forall a,b\in\genericspace$,
\begin{equation}
    \definition{\twovectorsaction[a][b]}{\mapdefinition{v}{\genericmetric[a][v]b-\genericmetric[b][v]a}},
\end{equation}
\noindent enabling us to write,
\begin{subequations}
\begin{gather}
    a_{-}\big(\composition{R(a)}{R(b)}\big)=-2\genericmetric[a][b]\twovectorsaction[a][b]\label{eqn:conformal-reflection-commutator}\\
    a_{+}\big(\composition{R(a)}{R(b)}\big)=\genericmetric[a][a]\genericmetric[b][b]\textup{id}+2\composition{\twovectorsaction[a][b]}{\twovectorsaction[a][b]}.\label{eqn:conformal-reflection-anticommutator}
\end{gather}
\end{subequations}
\noindent Since $a_{+}(A)+a_{-}(A)=A$, we find that the map $\twovectorsaction[a][b]$ controls binary products of conformal reflections. It is also antisymmetric, $\twovectorsaction[b][a]=-\twovectorsaction[a][b]$, and anti-self-\gadjoint. Furthermore, given \eqref{eqn:conformal-reflection-commutator}, and $\forall v\in\genericspace$,
\begin{subequations}
\begin{gather}
    a_{-}\big(\composition{R(a)}{\twovectorsaction[b][c]}\big)=\half\big(\twovectorsaction[R(a)(b)][c]+\twovectorsaction[b][R(a)(c)]\big)\\
    \genericmetric[b][c]\,a_{+}\big(\composition{R(a)}{\twovectorsaction[b][c]}\big)=\half\big(\genericmetric^{2}(a{,}c)R(b)-\genericmetric^{2}(a{,}b)R(c)-R(\twovectorsaction[a][b](c))+R(\twovectorsaction[a][c](b))\big),
\end{gather}
\end{subequations}
\noindent we see the $\set{R(a),t(b,c)}$ forms a generating set for the algebra of conformal reflections.

Therefore, we have derived a map with central importance to the conformal reflections, and whose image agrees with the image of \eqref{eqn:sothree-action-generic}. So far our derivation has only accounted for non-null vectors $a,b\in\genericspace$ with $\genericmetric[a][b]\neq0$, so,
\begin{equation}
    \twovectorsaction[a][b]=\frac{\composition{R(a)}{R(b)}-\composition{R(b)}{R(a)}}{-4\genericmetric[a][b]}.
\end{equation}
\noindent Let us extend this definition to the whole of $\genericspace$. Since $\twovectorsaction[a][b+\epsilon a]=\twovectorsaction[a][b]$, $\forall\epsilon\in\reals$, $\twovectorsaction[a][b]$ for non-null vectors is defined without the need for limits. To define $\twovectorsaction[a][b]$ when $b$ is null and $\genericmetric[a][b]=0$, by the non-degeneracy of $\genericmetric$ we may find a null $c\in\genericspace$ such that $\genericmetric[b][c]\neq0$, and use this pair to construct a pair of vectors $\set{p,n}$ such that,
\begin{subequations}
\begin{gather}
    \genericmetric[p][p]=-\genericmetric[n][n]>0\\
    \genericmetric[p][n]=0\\
    b=p+n\\
    c=p-n.
\end{gather}
\end{subequations}
\noindent Thus, we may define,
\begin{equation}
    t(a,b)=t(a,p)+t(a,n),
\end{equation}
\noindent by the bilinearity of $\twovectorsaction$. Similar arguments yield $\twovectorsaction[a][b]$ for $a$ and $b$ both null, regardless of the value of $\genericmetric[a][b]$. Thus, we have defined $\twovectorsaction[a][b]$ $\forall a,b\in\genericspace$ and completed our derivation of the $\sopq$-action on $\genericspace$; this was achieved in a coordinate-free, elementary way, without appealing to differential geometry or the theory of Lie groups\cite{varadarajan}.

\subsubsection{The Algebraic Form of \texorpdfstring{$\twovectorsaction[a][b]$}{t(a,b)}}

Having acquired $\twovectorsaction[a][b]$, the $\sopq$-action on $\genericspace$, $\forall a,b\in\genericspace$, we must consider how to implement it algebraically within an associative algebra. More precisely, we seek a third-order tensor $f(a,b,c)$ whose properties match those of $\twovectorsaction[a][b](c)$, so that a quotient of $\generictensoralgebra$ by the two-sided ideal $\forall a,b,c\in\genericspace$,
\begin{equation}
    \ideal{f(a,b,c)-\twovectorsaction[a][b](c)}
\end{equation}
\noindent yields the most general non-trivial algebra possible.

We first note that $\twovectorsaction[a][b](c)$ is antisymmetric in its first two arguments. The most general third-order tensor sharing this property is,
\begin{equation}\label{eqn:f-tensor-definition}
    f(a,b,c)=k_{1}\big(\tensor{(\tensor{a;b}-\tensor{b;a});c}\big)+k_{2}\big(\tensor{c;(\tensor{a;b}-\tensor{b;a})}\big)+k_{3}\big(\tensor{a;c;b}-\tensor{b;c;a}\big).
\end{equation}
\noindent We may constrain \eqref{eqn:f-tensor-definition} even further by considering  the commutator between two $\twovectorsaction$ maps, which is closed,
\begin{equation}\label{eqn:action-commutator}
    \composition{\twovectorsaction[a][b]}{\twovectorsaction[c][d]}-\composition{\twovectorsaction[c][d]}{\twovectorsaction[a][b]}=\twovectorsaction[\twovectorsaction[a][b](c)][d]+\twovectorsaction[c][\twovectorsaction[a][b](d)].
\end{equation}
\noindent From \eqref{eqn:action-commutator}, we see $\forall a,b,c,d,e\in\genericspace$,
\begin{equation}
    \twovectorsaction[\twovectorsaction[a][b](c)][d](e)+\twovectorsaction[c][\twovectorsaction[a][b](d)](e)=-\twovectorsaction[\twovectorsaction[c][d](a)][b](e)-\twovectorsaction[a][\twovectorsaction[c][d](b)](e),
\end{equation}
\noindent from which we require,
\begin{equation}\label{eqn:commutator-constraint}
    f(f(a,b,c),d,e)+f(c,f(a,b,d),e)+f(f(c,d,a),b,e)+f(a,f(c,d,b),e)=0.
\end{equation}
\noindent The only non-trivial solution to \eqref{eqn:commutator-constraint} is,
\begin{equation*}\label{eqn:f-tensor-constrained}
    f(a,b,c)=k_{1}\big(\tensor{(\tensor{a;b}-\tensor{b;a});c}-\tensor{c;(\tensor{a;b}-\tensor{b;a})}\big),
\end{equation*}
\noindent defining $f(a,b,c)$ up to an arbitrary scaling, which was expected.

\subsection{The Spinless Weak Clifford Algebra}\label{sec:spinless-clifford-algebra}

We now have everything we need to construct the \enquote{Spinless Weak Clifford Algebra} $\spinlesscliffordalgebra$: choosing $k_{1}=\half$ in \eqref{eqn:f-tensor-constrained}, we define $\forall a,b,c\in\genericspace$,
\begin{equation}\label{eqn:spinless-clifford-algebra}
    \spinlesscliffordalgebra\cong\frac{\generictensoralgebra}{\ideal{\tensor{(\tensorwedge{a}{b});c}-\tensor{c;(\tensorwedge{a}{b})}-\genericmetric[a][c]b+\genericmetric[b][c]a}},
\end{equation}
\noindent whose defining identity,
\begin{equation}\label{eqn:lie-algebra-action}
    \tensor{(\tensorwedge{a}{b});c}-\tensor{c;(\tensorwedge{a}{b})}=\genericmetric[a][c]b-\genericmetric[b][c]a,
\end{equation}
\noindent agrees with the Clifford algebra's $\sopq$-action \eqref{eqn:clifford-bivector-action} up to a scaling. We use the term \enquote{weak} Clifford algebra to contrast the \enquote{strong} Clifford algebra in the sense used in logic: the defining relationship of $\genericcliffordalgebra$ is stronger than the defining relationship of $\spinlesscliffordalgebra$.

The bivector $\tensorwedge{a}{b}$ necessarily appearing whole in \eqref{eqn:lie-algebra-action} demonstrates its significance to the properties of $\genericmetric$, and that $\twovectorsaction[a][b]$ is truly a bivector-action on a vector. We may capture this by defining,
\begin{equation}\label{eqn:bivector-action}
    \definition{\bivectoraction[\tensorwedge{a}{b}]}{\mapdefinition{c}{\tensor{(\tensorwedge{a}{b});c}-\tensor{c;(\tensorwedge{a}{b})}}}=\twovectorsaction[a][b].
\end{equation}
\noindent As the unique embedding of $\twovectorsaction[a][b]$ in $\spinlesscliffordalgebra$, we may consider the properties of $\bivectoraction[\tensorwedge{a}{b}]$ to be the natural extension of $\twovectorsaction[a][b]$ to $\spinlesscliffordalgebra$. In particular, we find that $\bivectoraction[\tensorwedge{a}{b}]$ is naturally a derivation\cite{bourbaki} on $\spinlesscliffordalgebra$, $\forall A,B\in\generictensoralgebra$,
\begin{equation}\label{eqn:bivector-action-derivation-property}
    \bivectoraction[\tensorwedge{a}{b}](\tensor{A;B})=\tensor{\big(\bivectoraction[\tensorwedge{a}{b}](A)\big);B}+\tensor{A;\big(\bivectoraction[\tensorwedge{a}{b}](B)\big)},
\end{equation} since,
\begin{equation}
    \tensor{(\tensorwedge{a}{b});\big(\tensor{A;B}\big)}-\tensor{\big(\tensor{A;B}\big);(\tensorwedge{a}{b})}=\tensor{\big(\tensor{(\tensorwedge{a}{b});A}-\tensor{A;(\tensorwedge{a}{b})}\big);B}+\tensor{A;\big(\tensor{(\tensorwedge{a}{b});B}-\tensor{B;(\tensorwedge{a}{b})}\big)}.
\end{equation}
\noindent This agrees with the standard prescription for a Lie algebra-action on a tensor product of representations, so that exponentiation yields a representation of a Lie group\cite{hall}. To ensure consistency with the structure of the $\spinalgebra{s}$, we shall extend $\bivectoraction$ to $T(\genericantisymmetrictensors{2})$ as an associative algebra-action, $\forall\alpha\in\reals,\,A,B\in T(\genericantisymmetrictensors{2})$,%
\begin{subequations}
\begin{gather}
    \definition{\bivectoraction[\alpha]}{\mapdefinition{A}{\alpha A}}\\
    \definition{\bivectoraction[\tensor{A;B}]}{\composition{\bivectoraction[A]}{\bivectoraction[B]}}\label{eqn:bivector-action-associative-property}.
\end{gather}
\end{subequations}

As with \eqref{eqn:clifford-bivector-action}, \eqref{eqn:lie-algebra-action} determines the Lie product of $\sopq$, which in $\spinlesscliffordalgebra$ is, $\forall a,b,c,d\in\genericspace$,
\begin{equation}\label{eqn:lie-product}
\begin{aligned}
    \tensor{(\tensorwedge{a}{b});(\tensorwedge{c}{d})}-\tensor{(\tensorwedge{c}{d});(\tensorwedge{a}{b})}=\genericmetric[a][c](\tensorwedge{b}{d})-\genericmetric[b][c](\tensorwedge{a}{d})-\genericmetric[a][d](\tensorwedge{b}{c})+\genericmetric[b][d](\tensorwedge{a}{c}),
\end{aligned}
\end{equation}
\noindent consistent with the standard Lie product\cite{weinberg-qft} with $\hat{J}^{\mu\nu}=\unitimaginary J^{\mu\nu}=\unitimaginary(\tensorwedge{e^{\mu}}{e^{\nu}})$, and $\hat{J}^{\mu\nu}$ the generators in the physics convention. This is also consistent with the spin generator convention $\hat{S}_{a}=\unitimaginary\generator[a]$ in \cite{bradshaw}. By properties \eqref{eqn:bivector-action-associative-property} and \eqref{eqn:bivector-action-derivation-property}, we may use \eqref{eqn:lie-product} to write the commutator of $\bivectoraction$ compactly,
\begin{equation}\label{eqn:bivector-action-lie-product}
    \composition{\bivectoraction[\tensorwedge{a}{b}]}{\bivectoraction[\tensorwedge{c}{d}]}-\composition{\bivectoraction[\tensorwedge{c}{d}]}{\bivectoraction[\tensorwedge{a}{b}]}=\bivectoraction[\bivectoraction[\tensorwedge{a}{b}](\tensorwedge{c}{d})],
\end{equation}
\noindent showing that $\bivectoraction$ is indeed an $\sopq$-action.

\subsection{The Spinless Weak Clifford Algebra for \texorpdfstring{$\threemetricspace$}{(E,delta)}}

Restricting our attention to the present problem, we consider the three-dimensional Euclidean space $\threemetricspace$ from earlier, and its spinless weak Clifford algebra $\spinlessthreecliffordalgebra$. Using an orthonormal basis $\set{e_{a}}$, we define,
\begin{equation}\label{eqn:spin-generator-bivector-map}
    \definition{\generator[p]}{\half\sothreesum*[r]{a,b}\sothreestructureconstants{a}{b}{p}\,\tensorwedge{e_{a}}{e_{b}}},
\end{equation}
\noindent which in $\spinlessthreecliffordalgebra$ turns \eqref{eqn:lie-product} into the standard Lie product of $\sothree$ \eqref{eqn:so3-lie-product}. Therefore, $\usothree\subset\spinlesscliffordalgebra$, and we identify,
\begin{equation}
    \domainrestriction{\bivectoraction}{T(\genericantisymmetrictensors{2})}=\sothreead,
\end{equation}
\noindent where $\sothreead$ was defined in \eqref{eqn:adjoint-action}. We note the inverse transformation of \eqref{eqn:spin-generator-bivector-map},
\begin{equation}\label{eqn:bivector-spin-generator-map}
    \tensorwedge{e_{a}}{e_{b}}=\sothreesum*[r]{p}\sothreestructureconstants{a}{b}{p}\generator[p],
\end{equation}
\noindent and recognise that it enables us to write the multipoles of the $\spinalgebra{s}$ in the language of bivectors, $\forall k\in\integers^{+}$,
\begin{equation}
    \multipole{k}\Big(\bigotimes_{j=1}^{k}\tensorwedge{e_{a_{j}}}{e_{b_{j}}}\Big)=\sothreesum*[l]{p_{1},\dots,p_{k}}\prod_{j=1}^{k}\sothreestructureconstants{a_{j}}{b_{j}}{p_{j}}\multipole{k}\Big(\bigotimes_{m=1}^{k}\generator[p_{m}]\Big),
\end{equation}
\noindent with,
\begin{equation}
    \sothreecasimirelement=\half\sothreesum*[r]{a,b}\tensor{(\tensorwedge{e_{a}}{e_{b}});(\tensorwedge{e_{a}}{e_{b}})}.
\end{equation}
\noindent Significantly, though $\usothree\subset\spinlessthreecliffordalgebra$, $\spinlessthreecliffordalgebra$ has no spin structure whatsoever. This makes it the ideal basic structure with which to realise the $\spinalgebra{s}$ and explore their geometric content.

\subsection{Measures of \texorpdfstring{$k$}{k}-Volumes}

\subsubsection{\texorpdfstring{$k$}{k}-Volumes in \texorpdfstring{$\spinlesscliffordalgebra$}{Cl-s-w(V,g)}}

In $\genericcliffordalgebra$, geometric measurements about its objects are conveyed by the scalar part\cite{doran-lasenby} $\langle\cdot\rangle$, for example: $|\langle(\tensorwedge{v_{1}}{\dots}{v_{k}})^{2}\rangle|$ is the square of a $k$-blade's $k$-volume; and $\langle(\tensorwedge{v_{1}}{\dots}{v_{k}})(\tensorwedge{w_{1}}{\dots}{w_{k}})\rangle$ describes the projected overlap of two $k$-blades. In $\usothree\subset\spinlessthreecliffordalgebra$, there is a similar notion.

Recall from section \ref{sec:spin-algebras}, that all elements $A_{0}\in\usothree$ for which $\sothreead[\sothreecasimirelement](A_{0})=0$ are $\reals[\sothreecasimirelement]$-linear combinations of the monopole $\multipoletensor{}$. Thus, given an element $A\in\usothree$ on which $\sothreead[\sothreecasimirelement]$ has minimal polynomial $m(x)$, we define its \enquote{Monopole Part} $\mon[A]$,
\begin{equation}
    \definition{\mon[A]}{\begin{cases}
        \frac{n(\sothreead[\sothreecasimirelement])}{n(0)}(A) & m(x)=xn(x)\\
        0 & \textup{otherwise}.
    \end{cases}}
\end{equation}
\noindent We may use $\bivectoraction$ to identify geometrically meaningful objects within $\spinlesscliffordalgebra$ with objects forming simple $\bivectoraction[\usothree]$-modules; this is consistent with how the multipole tensors were identified\cite{bradshaw}. The $k$-blades are such modules, demonstrating their significance even in $\spinlesscliffordalgebra$.

\subsubsection{\texorpdfstring{$k$}{k}-Volumes in \texorpdfstring{$\spinlessthreecliffordalgebra$}{Cl-s-w(E,delta)}}

In $\spinlessthreecliffordalgebra$, we wish to capture the sizes of bivectors and trivectors in some natural way. Accordingly, the invariant tensors at second-order,
\begin{equation}
    \mon[\tensor{\generator[p];\generator[q]}]=\frac{1}{3}\delta_{pq}\sothreecasimirelement
\end{equation}
\noindent and third-order,
\begin{equation}
    \mon[\tensor{\generator[p];\generator[q];\generator[r]}]=\frac{1}{6}\sothreestructureconstants{p}{q}{r}\sothreecasimirelement
\end{equation}
\noindent in $\usothree$ are of particular interest, where we have used that $\multipoletensor{}=1$. In bivector form, these are respectively,
\begin{equation}
    \mon[\tensor{(\tensorwedge{a}{b});(\tensorwedge{c}{d})}]=\frac{1}{3}\big(\threemetric[a][c]\threemetric[b][d]-\threemetric[a][d]\threemetric[b][c]\big)\sothreecasimirelement,
\end{equation}
\noindent and,
\begin{equation}
    \mon[\tensor{(\tensorwedge{a}{b});(\tensorwedge{c}{d});(\tensorwedge{e}{f})}]=\frac{1}{6}\begin{aligned}[t]
        \Big(&\threemetric[a][c]\big(\threemetric[b][e]\threemetric[d][f]-\threemetric[b][f]\threemetric[d][e]\big)\\
        -&\threemetric[a][d]\big(\threemetric[b][e]\threemetric[c][f]-\threemetric[b][f]\threemetric[c][e]\big)\\
        -&\threemetric[a][e]\big(\threemetric[b][c]\threemetric[d][f]-\threemetric[b][d]\threemetric[c][f]\big)\\
        +&\threemetric[a][f]\big(\threemetric[b][c]\threemetric[d][e]-\threemetric[b][d]\threemetric[c][e]\big)\Big)\sothreecasimirelement
    \end{aligned}
\end{equation}
\noindent Since these objects are invariant under $\bivectoraction$, they are invariant under the action of $\specialorthogonalgroup{3}$, and we may use them to extend the metric $\threemetric$ of $\threemetricspace$ to a metric $\antimetric$ on the space of all antisymmetric tensors\cite{bourbaki} $\threeantisymmetrictensors{}$, $\forall\alpha,\beta\in\reals$, $\forall a,b,c,d,e,f\in\threespace$,
\begin{subequations}
\begin{gather}
    \definition{\antimetric[\alpha][\beta]}{\alpha\beta}\\
    \definition{\antimetric[a][b]}{\threemetric[a][b]}\\
    \begin{aligned}
        \definition{\antimetric[\tensorwedge{a}{b}][\,\tensorwedge{c}{d}]&}{\mon[\tensor{(\tensorwedge{a}{b});(\tensorwedge{c}{d})}]}\\
        &=\frac{1}{3}\cbmetric[\tensorwedge{a}{b}][\tensorwedge{c}{d}]\sothreecasimirelement
    \end{aligned}\\
    \begin{aligned}
        \definition{\antimetric[\tensorwedge{a}{b}{c}][\,\tensorwedge{d}{e}{f}]&}{\begin{aligned}[t]
            &\mon[\tensor{(\tensorwedge{a}{b});(\tensorwedge{c}{d});(\tensorwedge{e}{f})}]\\
            &-\mon[\tensor{(\tensorwedge{a}{b});(\tensorwedge{c}{e});(\tensorwedge{d}{f})}]\\
            &+\mon[\tensor{(\tensorwedge{a}{b});(\tensorwedge{c}{f});(\tensorwedge{d}{e})}]
        \end{aligned}}\\
        &=-\frac{1}{3}\cbmetric[\tensorwedge{a}{b}{c}][\tensorwedge{d}{e}{f}]\sothreecasimirelement,
    \end{aligned}
\end{gather}
\end{subequations}
\noindent with all other combinations zero, and $\cbmetric$ is the usual Cauchy-Binet metric\cite{bourbaki} for $\threemetricspace$, $\forall k\in\integers^{+}$,
\begin{equation}
    \definition{\cbmetric[\bigwedge_{j=1}^{n}a_{j}][\bigwedge_{k=1}^{n}b_{k}]}{\det(\threemetric[a_{j}][b_{k}])}.
\end{equation}
\noindent Note that $\antimetric$ is not (yet) scalar-valued,
\begin{equation}
    \mapdeclaration{\antimetric}{\threeantisymmetrictensors{}}{\reals[\sothreecasimirelement]}.
\end{equation}

%% file: results.tex
\section{Results}\label{sec:results}

\subsection{The Spin-\texorpdfstring{$s\neq0$}{s=/=0} Weak Clifford algebras}

Using the spinless weak Clifford algebra for Euclidean three-space $\spinlessthreecliffordalgebra$ of \eqref{eqn:spinless-clifford-algebra}, we may define the \enquote{Spin-$s$ Weak Clifford Algebra} $\spinscliffordalgebra{s}$ for spin-$s\neq0$,
\begin{equation}\label{eqn:spin-s-weak-clifford-definition}
    \spinscliffordalgebra{s}\cong\frac{\spinlessthreecliffordalgebra}{\ideal{\image{\multipole{2s+1}}}}.
\end{equation}
\noindent Since $\usothree\subset\spinlessthreecliffordalgebra$, this is equivalent to embedding the structure of the $\spinalgebra{s}$ within $\spinlessthreecliffordalgebra$. Within this algebra, we may positively identify the $\spinalgebra{s}$ as algebras of bivectors in general, whose action on $\threespace$ respects its Euclidean geometry $\threemetric$. Thus, we have finally explicated the structure of the $\spinalgebra{s}$ in both geometric and algebraic terms. However, the embedding of their spin structures also has an impact on the geometry itself. 

As before, the quotient \eqref{eqn:spin-s-weak-clifford-definition} entails,
\begin{equation}\label{eqn:casimir-element-weak-clifford}
    \sothreecasimirelement=-s(s+1),
\end{equation}
\noindent within $\spinscliffordalgebra{s}$. This ensures that the metric $\antimetric$ becomes scalar valued for bivectors and trivectors,
\begin{subequations}
\begin{gather}
    \antimetric[\tensorwedge{a}{b}][\,\tensorwedge{a}{b}]=-\frac{s(s+1)}{3}\cbmetric[\tensorwedge{a}{b}][\tensorwedge{a}{b}]\\
    \antimetric[\tensorwedge{a}{b}{c}][\,\tensorwedge{a}{b}{c}]=\frac{s(s+1)}{3}\cbmetric[\tensorwedge{a}{b}{c}][\tensorwedge{a}{b}{c}],
\end{gather}
\end{subequations}
\noindent and naturally \textit{spin dependent}. The metric on vectors and on scalars remains spin independent. Besides this feature, the values of the metric $\antimetric$ are quite different from those of the usual Clifford algebra on $\threemetricspace$. It has more consistency with $\threeconsistentcliffordalgebra$, for example the bivectors have the same size and sign, and the trivector has the same sign. However, the vectors and trivector are too large by a factor of $2$, and the vectors are also positive-definite. There is freedom in scaling $\antimetric$ in these sectors for consistency, but the author can see no mathematical reason for doing so at time of writing.

Despite this change to square norms, the algebraic structure imparted on $\spinscliffordalgebra{s}$ does not affect the action of $\bivectoraction[\tensorwedge{a}{b}]$ nor the rotational behaviour of $\threespace\subset\spinscliffordalgebra{s}$, since \eqref{eqn:spin-s-weak-clifford-definition} constrains only totally symmetric combinations of bivectors. However, the structure of $\spinscliffordalgebra{s}$ is significantly affected by the spin structure from $\spinalgebra{s}$. Recall that $\spinalgebra{s}$ is defined completely from $\usothree$ by $\image{\multipole{2s+1}}=\set{0}$. Taking spin-$\half$ as an example, this is equivalent to a series of tensor identities, $\forall a,b\in\set{1,2,3}$,
\begin{equation}
    \half(\tensor{\generator[a];\generator[b]}+\tensor{\generator[b];\generator[a]})+\frac{1}{4}\delta_{ab}=0.
\end{equation}
\noindent In the language of bivectors this condition is equivalent to $\forall a,b,c,d\in\threespace$,
\begin{equation}\label{eqn:quadrupole-bivector-identity}
    \half\big(\tensor{(\tensorwedge{a}{b});(\tensorwedge{c}{d})}+\tensor{(\tensorwedge{c}{d});(\tensorwedge{a}{b})}\big)=\antimetric[\tensorwedge{a}{b}][\,\tensorwedge{c}{d}].
\end{equation}
\noindent Recognising $\antimetric[\tensorwedge{a}{b}][\,\tensorwedge{c}{d}]$ as a scalar, we may break up each bivector on the left-hand side according to \eqref{eqn:antisymmetric-tensors}, revealing \eqref{eqn:quadrupole-bivector-identity} to be an constraint on fourth-order tensors in $\spinscliffordalgebra{1/2}$. For spin-$s$, these identities constrain order $2(2s+1)$ tensors. Interpreting $\threespace$ as physical Euclidean space, these embeddings of the spin structures of $\spinalgebra{s}$ within $\spinlessthreecliffordalgebra$ constitute non-commutative geometries for $\threespace$, in the sense of non-commuting position observables\cite{szabo,aschieri,frob}.

\subsection{The Spin-\texorpdfstring{$0$}{0} Weak Clifford algebra}

The case of the spin-$0$ algebra is an edge-case requiring separate treatment. $\spinalgebra{0}$ contains only the monopole $\multipoletensor{}=1$, and is defined by $\image{\multipole{1}}=\set{0}$, which entails $\sothreecasimirelement=0$. In the bivector language, this means that $\forall a,b\in\threespace$,
\begin{equation}
    \tensorwedge{a}{b}=0.
\end{equation}
\noindent Trying to apply this identity to $\spinlessthreecliffordalgebra$ as in \eqref{eqn:spin-s-weak-clifford-definition} results in the trivial algebra $\reals$. This is because in \eqref{eqn:spinless-clifford-algebra} we associate $\big[\threeantisymmetrictensors{2},\threespace\big]$ with the whole of $\threespace$, so quotienting all bivectors in $\spinlessthreecliffordalgebra$ sets all vectors in the algebra to $0$. Really, the identity \eqref{eqn:lie-algebra-action} is only reasonable when the bivectors are non-zero in the algebra, otherwise we seek an action of zero mapping any vector to any other. To avoid this, we must impose the structure of $\spinalgebra{0}$ directly on $\threetensoralgebra$,
\begin{equation}
    \spinscliffordalgebra{0}\cong\frac{\threetensoralgebra}{\image{\multipole{1}}}.
\end{equation}
\noindent This algebra is unique amongst the spin-$s$ weak Clifford algebras: $\image{\multipole{1}}=\set{0}$ implies that $\forall a,b\in\threespace$,
\begin{equation}
    \tensor{a;b}=\tensor{b;a},
\end{equation}
\noindent so $\spinscliffordalgebra{0}\cong\textup{Sym}(\threespace)$ is commutative. In fact, a spin-$s$ weak Clifford algebra is commutative iff it has a spin-$0$ structure. Additionally, $\sothreecasimirelement=0$ implies that,
\begin{subequations}
\begin{gather}
    \antimetric[\tensorwedge{a}{b}][\,\tensorwedge{a}{b}]=0\\
    \antimetric[\tensorwedge{a}{b}{c}][\,\tensorwedge{a}{b}{c}]=0,
\end{gather}
\end{subequations}
\noindent which is consistent with our expectations from the other spin-$s$ weak Clifford algebras, and the commutative nature of $\spinscliffordalgebra{0}$.

%% file: discussion.tex
\section{Discussion}\label{sec:discussion}

Interpreting the meaning of the spin-$s$ weak Clifford algebras depends heavily on our interpretation of the Euclidean three-space $\threespace$. The simplest and most relevant view of $\threespace$, is as the non-relativistic configuration space for a point-like particle. We may then interpret each vector as a position in three-space, or as the underlying algebraic object for a position operator in a quantum mechanical system. In this setting, we see that each spin-$s$ weak Clifford algebra describes a different algebra of position variables according to its spin structure: the spin-$0$ weak Clifford algebra is commutative and corresponds to the usual position operator algebra in quantum mechanics; and the higher spin algebras are all non-commutative.

In the sense of non-commuting position operators, we see a correspondence between the spin-$s$ weak Clifford algebras, and non-commutative geometries whose structures are determined by their spin. Though the meaning of the spin dependence of area and volume is unclear, especially when identically zero, these phenomena further indicate that the spin structure affects the geometry of, or perhaps experienced by, the system. These non-commutative geometries are, in general, much weaker than those common to the literature\cite{szabo,aschieri,frob}, which typically place the position operators into a Heisenberg-like\cite{schempp,wallach} algebra.

From these observations, we see the $\spinscliffordalgebra{s}$ as a new way to incorporate spin into quantum mechanical theories: directly as certain non-commutative algebras of position (and perhaps, by symmetry, momentum) operators. It is viable to extend the $\spinscliffordalgebra{s}$ to such a more phenomenologically complete model, since they contain the totally symmetric tensors, which are essential to algebraically perform dynamical (symplectic) transformations\cite{fulton-harris,crumeyrolle}. Aside from these considerations, $\spinscliffordalgebra{1/2}$ and $\spinscliffordalgebra{1}$ are also weak enough that the Euclidean Clifford and Duffin-Kemmer-Petiau\cite{fischbach,helmstetter,micali} algebras respectively may be derived from them. Thus, the $\spinscliffordalgebra{s}$ may form the basis for a generalised spin-$s$ theory of such algebras. Furthermore, relativistic versions of this formalism may prove useful in the construction of theories of quantum gravity which incorporate both non-commutative geometry and spin. 

With our interpretation of the $\spinscliffordalgebra{s}$ laid out, it is instructive to compare it against other arbitrary spin models. The most relevant such comparison is with the standard tensor product of center of mass and \enquote{internal} spin degrees of freedom\cite{weyl}. An immediate similarity is that both models include the spin algebra $\spinalgebra{s}$ as a subalgebra, originating from their spin structures. An immediate difference is that the traditional model incorporates the Heisenberg algebra\cite{schempp,wallach}, and therefore the notion of momentum, whereas the spin-$s$ weak Clifford algebras do not. However, the most significant difference is that in the traditional model the position and spin degrees of freedom are commuting, and thus independent of each other; they are not in $\spinscliffordalgebra{s}$ by construction. This implies that there is phenomenology between position and spin in a dynamical model containing $\spinscliffordalgebra{s}$ which the tensor product model cannot describe. The tensor product model should however be recoverable within this richer formalism as an approximation in some suitable setting.

Another standard approach to higher spin in non-relativistic physics is to consider a subspace of $\tensorpower{\threecliffordalgebra}{k}$ with the appropriate spin structure\cite{sommen}. However, due to the strength of the algebra, it lacks totally symmetric tensors, and so cannot easily form part of a model which algebraically encodes symplectic transformations. Such algebras also have interpretational issues regarding the underlying substructure of $\tensorpower{\threecliffordalgebra}{k}$ when applying it to fundamental particles; $\spinscliffordalgebra{s}$ does not suffer from this. Beyond the realm of non-relativistic physics are the Bargmann-Wigner\cite{bargmann} and Joos-Weinberg\cite{jefferey,weinberg-spin} equations. Since the former equations do not have definite spin in general\cite{jaroszewicz}, we shall focus on the latter. The $\gamma^{\mu_{1}\dots\mu_{2s}}$ for a particle of spin-$s$ in the Joos-Weinberg equation are comprised of objects which bear close, but not exact, resemblance to the multipole tensors of order $s$\cite{bradshaw}, revealing a link to the spin structure of the theory. Much like the tensor product model however, the spin sectors of the Joos-Weinberg equations and their center of mass sectors commute, as do the position operators. In this way the comparisons made between $\spinscliffordalgebra{s}$ and the tensor product model are valid for the Joos-Weinberg equations also.

%% file: conclusion.tex
\section{Conclusion}

In this paper, we demonstrated the incompatibility of the Clifford algebra with arbitrary non-relativistic spin structures, and defined a family of generalised Clifford-like algebras which support arbitrary spin structures. To do this, we presented a novel algebraic derivation for the structure of the Lie algebra $\sopq$, without the need to appeal to the theory of differential geometry or Lie groups. We also defined an even more general Clifford-like algebra with no spin structure at all, which underpins the structure of the arbitrary spin Clifford-like algebras. In so doing, we explicated the geometric structure of the spin algebras, and showed that they each define a unique non-commutative geometry on Euclidean three-space. We found that this structure induces a spin-dependence on the measured notions of area and volume, and compared these new arbitrary spin algebras with existing models of arbitrary spin. Some applications and avenues for further enquiry were discussed.

%% file: acknowledgements.tex
\section{Acknowledgements}

The author would like to thank B. J. Hiley, M. Hajtanian, and D. Nellist for their insightful conversations and support.

%% file: references.bib
@article{fischbach,
	title = {The Lie algebra s o (N) and the Duffin‐Kemmer‐Petiau ring},
	volume = {15},
	issn = {0022-2488},
	url = {https://doi.org/10.1063/1.1666504},
	doi = {10.1063/1.1666504},
	pages = {60--64},
	number = {1},
	journaltitle = {Journal of Mathematical Physics},
	shortjournal = {Journal of Mathematical Physics},
	author = {Fischbach, Ephraim and Louck, James D. and Nieto, Michael Martin and Scott, C. K.},
	urldate = {2023-05-30},
	date = {2003-11-04},
}

@article{helmstetter,
	title = {About the Structure of Meson Algebras},
	volume = {20},
	issn = {1661-4909},
	url = {https://doi.org/10.1007/s00006-010-0213-0},
	doi = {10.1007/s00006-010-0213-0},
	pages = {617--629},
	number = {3},
	journaltitle = {Advances in Applied Clifford Algebras},
	shortjournal = {Adv. Appl. Clifford Algebras},
	author = {Helmstetter, Jacques and Micali, Artibano},
	urldate = {2022-03-23},
	date = {2010-10-01},
	langid = {english},
}

@article{micali,
	title = {On Meson Algebras},
	volume = {18},
	issn = {1661-4909},
	url = {https://doi.org/10.1007/s00006-008-0118-3},
	doi = {10.1007/s00006-008-0118-3},
	pages = {875--889},
	number = {3},
	journaltitle = {Advances in Applied Clifford Algebras},
	shortjournal = {{AACA}},
	author = {Micali, Artibano and Rachidi, Mustapha},
	urldate = {2022-03-23},
	date = {2008-09-01},
	langid = {english},
}

@misc{frob,
	title = {Non-commutative Geometry from Perturbative Quantum Gravity in de Sitter spacetime},
	url = {http://arxiv.org/abs/2305.01517},
	doi = {10.48550/arXiv.2305.01517},
	number = {{arXiv}:2305.01517},
	publisher = {{arXiv}},
	author = {Fröb, Markus and Lima, William C. C. and Much, Albert and Papadopoulos, Kyriakos},
	urldate = {2023-05-29},
	date = {2023-04-29},
	eprinttype = {arxiv},
	eprint = {2305.01517 [gr-qc]},
}

@book{aschieri,
	location = {Berlin, Heidelberg},
	title = {Noncommutative Spacetimes: Symmetries in Noncommutative Geometry and Field Theory},
	volume = {774},
	isbn = {9783540897927 9783540897934},
	url = {https://link.springer.com/10.1007/978-3-540-89793-4},
	series = {Lecture Notes in Physics},
	shorttitle = {Noncommutative Spacetimes},
	publisher = {Springer},
	author = {Aschieri, Paolo and Dimitrijevic, Marija and Kulish, Petr and Lizzi, Fedele and Wess, Julius},
	urldate = {2023-05-29},
	date = {2009},
	langid = {english},
	doi = {10.1007/978-3-540-89793-4},
}

@book{varadarajan,
	location = {New York, {NY}},
	title = {Lie Groups, Lie Algebras, and Their Representations},
	volume = {102},
	isbn = {9781461270164 9781461211266},
	url = {http://link.springer.com/10.1007/978-1-4612-1126-6},
	series = {Graduate Texts in Mathematics},
	publisher = {Springer},
	author = {Varadarajan, V. S.},
	urldate = {2023-05-29},
	date = {1984},
	doi = {10.1007/978-1-4612-1126-6},
}

@book{hall,
	location = {Cham},
	title = {Lie Groups, Lie Algebras, and Representations: An Elementary Introduction},
	volume = {222},
	isbn = {9783319134666 9783319134673},
	url = {https://link.springer.com/10.1007/978-3-319-13467-3},
	series = {Graduate Texts in Mathematics},
	shorttitle = {Lie Groups, Lie Algebras, and Representations},
	publisher = {Springer International Publishing},
	author = {Hall, Brian C.},
	urldate = {2023-05-29},
	date = {2015},
	langid = {english},
	doi = {10.1007/978-3-319-13467-3},
}

@book{weinberg-qft,
	location = {Cambridge},
	title = {The Quantum Theory of Fields: Volume 1: Foundations},
	volume = {1},
	isbn = {9780521670531},
	url = {https://www.cambridge.org/core/books/quantum-theory-of-fields/22986119910BF6A2EFE42684801A3BDF},
	shorttitle = {The Quantum Theory of Fields},
	publisher = {Cambridge University Press},
	author = {Weinberg, Steven},
	urldate = {2023-05-27},
	date = {1995},
	doi = {10.1017/CBO9781139644167},
}

@article{jaroszewicz,
	title = {Geometry of spacetime propagation of spinning particles},
	volume = {216},
	issn = {0003-4916},
	url = {https://www.sciencedirect.com/science/article/pii/000349169290176M},
	doi = {10.1016/0003-4916(92)90176-M},
	pages = {226--267},
	number = {2},
	journaltitle = {Annals of Physics},
	shortjournal = {Annals of Physics},
	author = {Jaroszewicz, T and Kurzepa, P. S},
	urldate = {2023-05-26},
	date = {1992-06-01},
	langid = {english},
}

@article{jefferey,
	title = {Component Minimization of the Bargmann-Wigner Wavefunction},
	volume = {31},
	pages = {137--149},
	number = {2},
	journaltitle = {Australian Journal of Physics},
	author = {Jefferey, E.A.},
	date = {1978},
	note = {{INIS} Reference Number: 10438537},
}

@article{szabo,
	title = {Quantum Field Theory on Noncommutative Spaces},
	volume = {378},
	issn = {03701573},
	url = {http://arxiv.org/abs/hep-th/0109162},
	doi = {10.1016/S0370-1573(03)00059-0},
	pages = {207--299},
	number = {4},
	journaltitle = {Physics Reports},
	shortjournal = {Physics Reports},
	author = {Szabo, Richard J.},
	urldate = {2023-02-22},
	date = {2003-05},
	eprinttype = {arxiv},
	eprint = {hep-th/0109162},
}

@inproceedings{sommen,
	title = {Clifford Tensor Calculus},
	pages = {423--436},
	booktitle = {Proc. 22th Conf. on Diff. Geom. Meth. in Theor. Phys.},
	author = {Sommen, Franciscus},
	date = {1993},
}

@article{weinberg-spin,
	title = {Feynman Rules for Any Spin},
	volume = {133},
	url = {https://link.aps.org/doi/10.1103/PhysRev.133.B1318},
	doi = {10.1103/PhysRev.133.B1318},
	pages = {B1318--B1332},
	number = {5},
	journaltitle = {Physical Review},
	shortjournal = {Phys. Rev.},
	author = {Weinberg, Steven},
	urldate = {2022-11-28},
	date = {1964-03-09},
}

@article{bargmann,
	title = {Group Theoretical Discussion of Relativistic Wave Equations},
	volume = {34},
	issn = {0027-8424},
	url = {https://www.ncbi.nlm.nih.gov/pmc/articles/PMC1079095/},
	pages = {211--223},
	number = {5},
	journaltitle = {Proceedings of the National Academy of Sciences of the United States of America},
	shortjournal = {Proc Natl Acad Sci U S A},
	author = {Bargmann, V. and Wigner, E. P.},
	urldate = {2022-11-28},
	date = {1948-05},
	pmid = {16578292},
	pmcid = {PMC1079095},
}

@misc{bradshaw,
	title = {An Algebraic Theory of Non-Relativistic Spin},
	url = {http://arxiv.org/abs/2207.02351},
	doi = {10.48550/arXiv.2207.02351},
	number = {{arXiv}:2207.02351},
	publisher = {{arXiv}},
	author = {Bradshaw, Peter T. J.},
	urldate = {2022-10-07},
	date = {2022-10-06},
	eprinttype = {arxiv},
	eprint = {2207.02351 [math-ph, physics:quant-ph]},
}

@book{fulton-harris,
	title = {Representation Theory: A First Course},
	isbn = {9780387974958},
	shorttitle = {Representation Theory},
	pagetotal = {574},
	publisher = {Springer},
	author = {Fulton, William and Harris, Joe},
	date = {1991},
	langid = {english},
	note = {Google-Books-{ID}: 6GUH8ARxhp8C},
}

@book{bourbaki,
	title = {Algebra I: Chapters 1-3},
	isbn = {9783540642435},
	shorttitle = {Algebra I},
	pagetotal = {750},
	publisher = {Springer},
	author = {Bourbaki, N.},
	date = {1998-08-03},
	langid = {english},
	note = {Google-Books-{ID}: {STS}9aZ6F204C},
}

@book{humphreys,
	title = {Introduction to Lie Algebras and Representation Theory},
	isbn = {9783540900535},
	pagetotal = {200},
	publisher = {Springer},
	author = {Humphreys, James E.},
	date = {1972},
	langid = {english},
	note = {Google-Books-{ID}: {TiUlAQAAIAAJ}},
}

@book{weyl,
	title = {The Theory of Groups and Quantum Mechanics},
	isbn = {9780486602691},
	pagetotal = {468},
	publisher = {Courier Corporation},
	author = {Weyl, Hermann},
	date = {1950-01-01},
	langid = {english},
	note = {Google-Books-{ID}: {jQbEcDDqGb}8C},
}

@incollection{schempp,
	location = {Dordrecht},
	title = {Radar Ambiguity Functions, Nilpotent Harmonic Analysis, and Holomorphic Theta Series},
	isbn = {9789401097871},
	url = {https://doi.org/10.1007/978-94-010-9787-1_6},
	series = {Mathematics and Its Applications},
	pages = {217--260},
	booktitle = {Special Functions: Group Theoretical Aspects and Applications},
	publisher = {Springer Netherlands},
	author = {Schempp, Walter},
	editor = {Askey, R. A. and Koornwinder, T. H. and Schempp, W.},
	urldate = {2020-08-26},
	date = {1984},
	langid = {english},
	doi = {10.1007/978-94-010-9787-1_6},
}

@book{wallach,
	location = {Brookline, Mass.},
	title = {Symplectic geometry and Fourier analysis},
	isbn = {9780915692156},
	url = {https://catalog.hathitrust.org/Record/003496599},
	series = {Lie groups : history, frontiers and applications ;5},
	pagetotal = {xvii, 436 p.},
	publisher = {Math Sci Press},
	author = {Wallach, Nolan R.},
	urldate = {2020-08-26},
	date = {1977},
}

@book{crumeyrolle,
	title = {Orthogonal and Symplectic Clifford Algebras: Spinor Structures},
	isbn = {9780792305415},
	url = {https://www.springer.com/gp/book/9780792305415},
	series = {Mathematics and Its Applications},
	shorttitle = {Orthogonal and Symplectic Clifford Algebras},
	publisher = {Springer Netherlands},
	author = {Crumeyrolle, A.},
	urldate = {2020-08-26},
	date = {1990},
	langid = {english},
	doi = {10.1007/978-94-015-7877-6},
}

@book{doran-lasenby,
	location = {Cambridge},
	title = {Geometric Algebra for Physicists},
	isbn = {9780521715959},
	url = {https://www.cambridge.org/core/books/geometric-algebra-for-physicists/FB8D3ACB76AB3AB10BA7F27505925091},
	publisher = {Cambridge University Press},
	author = {Doran, Chris and Lasenby, Anthony},
	urldate = {2020-08-26},
	date = {2003},
	doi = {10.1017/CBO9780511807497},
}
